\pdfoutput=1
\documentclass[12pt,a4paper]{article}

\usepackage[section]{placeins}
\usepackage{subcaption}
\usepackage{makecell}
\usepackage{multirow}
\usepackage{comment}
\usepackage{overpic}
\usepackage{tikz-feynman}

\usepackage{ifthen} 
\newboolean{pdflatex}
\setboolean{pdflatex}{true} 

\newboolean{articletitles}
\setboolean{articletitles}{true} 

\newboolean{uprightparticles}
\setboolean{uprightparticles}{false} 
\usepackage[figuresright]{rotating}

\def\paperauthors{LHCb collaboration} 
\def\paperasciititle{Observation of the doubly charmed baryon Omega_cc^+} 
\def\papertitle{Observation of the doubly charmed baryon $\varOmega^+_{cc}$}
\def\paperkeywords{{HEP}, {LHCb}} 
\def\papercopyright{\the\year\ CERN for the benefit of the LHCb collaboration} 
\def\paperlicence{CC BY 4.0 licence}
\def\paperlicenceurl{https://creativecommons.org/licenses/by/4.0/}

\usepackage[top=1in, bottom=1.25in, left=1in, right=1in]{geometry}

\usepackage{microtype}
\usepackage{lineno}  
\usepackage{xspace} 
\usepackage{caption} 

\usepackage{graphicx}  
\usepackage{color}
\usepackage{colortbl}
\graphicspath{{./figs/}} 

\usepackage{amsmath} 
\usepackage{amssymb}
\usepackage{amsfonts}
\usepackage{upgreek} 

\newcommand*\patchAmsMathEnvironmentForLineno[1]{%
\expandafter\let\csname old#1\expandafter\endcsname\csname #1\endcsname
\expandafter\let\csname oldend#1\expandafter\endcsname\csname
end#1\endcsname
 \renewenvironment{#1}%
   {\linenomath\csname old#1\endcsname}%
   {\csname oldend#1\endcsname\endlinenomath}%
}
\newcommand*\patchBothAmsMathEnvironmentsForLineno[1]{%
  \patchAmsMathEnvironmentForLineno{#1}%
  \patchAmsMathEnvironmentForLineno{#1*}%
}
\AtBeginDocument{%
\patchBothAmsMathEnvironmentsForLineno{equation}%
\patchBothAmsMathEnvironmentsForLineno{align}%
\patchBothAmsMathEnvironmentsForLineno{flalign}%
\patchBothAmsMathEnvironmentsForLineno{alignat}%
\patchBothAmsMathEnvironmentsForLineno{gather}%
\patchBothAmsMathEnvironmentsForLineno{multline}%
\patchBothAmsMathEnvironmentsForLineno{eqnarray}%
}

\usepackage{hyperxmp}

\usepackage[pdftex,
            pdfauthor={\paperauthors},
            pdftitle={\paperasciititle},
            pdfkeywords={\paperkeywords},
            pdfcopyright={Copyright (C) \papercopyright},
            pdflicenseurl={\paperlicenceurl}]{hyperref}

\usepackage[colorinlistoftodos,textsize=scriptsize]{todonotes}

\usepackage[bottom,flushmargin,hang,multiple]{footmisc}

\usepackage[all]{hypcap} 

\usepackage{mwe} 

\usepackage{longtable} 

\usepackage{booktabs}

\usepackage{xspace} 
\usepackage{upgreek}

\def\lhcb   {\mbox{LHCb}\xspace}

\def\MagUp {\mbox{\em Mag\kern -0.05em Up}\xspace}

\ifthenelse{\boolean{uprightparticles}}%
{

 \def\Pmu         {\ensuremath{\upmu}\xspace}

 \def\Ppi         {\ensuremath{\uppi}\xspace}

 \def\Ppsi        {\ensuremath{\uppsi}\xspace}

 \def\PDelta      {\ensuremath{\Delta}\xspace}                 
 \def\PXi         {\ensuremath{\Xi}\xspace}                 
 \def\PLambda     {\ensuremath{\Lambda}\xspace}                 
 \def\PSigma      {\ensuremath{\Sigma}\xspace}                 
 \def\POmega      {\ensuremath{\Omega}\xspace}                 
 \def\PUpsilon    {\ensuremath{\Upsilon}\xspace}
 \let\oldPi\Pi
 \def\PPi         {\ensuremath{\oldPi}\xspace}

 \def\PB      {\ensuremath{\mathrm{B}}\xspace}                 
 \def\PD      {\ensuremath{\mathrm{D}}\xspace}                 
 \def\PJ      {\ensuremath{\mathrm{J}}\xspace}                 
 \def\PK      {\ensuremath{\mathrm{K}}\xspace}                 
 \def\Pb      {\ensuremath{\mathrm{b}}\xspace}                 
 \def\Pc      {\ensuremath{\mathrm{c}}\xspace}

 \def\Pp      {\ensuremath{\mathrm{p}}\xspace}                 

 \def\Ps      {\ensuremath{\mathrm{s}}\xspace}

 \def\thebaroffset{0.0em}
}
{

 \def\Pmu         {\ensuremath{\mu}\xspace}

 \def\Ppi         {\ensuremath{\pi}\xspace}

 \def\Ppsi        {\ensuremath{\psi}\xspace}                 
                  
 \mathchardef\PDelta="7101
 \mathchardef\PXi="7104
 \mathchardef\PLambda="7103
 \mathchardef\PSigma="7106
 \mathchardef\POmega="710A
 \mathchardef\PUpsilon="7107
 \mathchardef\PPi="7105
 \def\PB      {\ensuremath{B}\xspace}                 
 \def\PD      {\ensuremath{D}\xspace}                 
 \def\PJ      {\ensuremath{J}\xspace}                 
 \def\PK      {\ensuremath{K}\xspace}                 
 \def\Pb      {\ensuremath{b}\xspace}                 
 \def\Pc      {\ensuremath{c}\xspace}

 \def\Pp      {\ensuremath{p}\xspace}                 

 \def\Ps      {\ensuremath{s}\xspace}

 \def\thebaroffset{0.18em}
}
\newcommand{\offsetoverline}[2][\thebaroffset]{\kern #1\overline{\kern -#1 #2}}%

\makeatletter
\ifcase \@ptsize \relax
  \newcommand{\miniscule}{\@setfontsize\miniscule{4}{5}}
\or
  \newcommand{\miniscule}{\@setfontsize\miniscule{5}{6}}
\or
  \newcommand{\miniscule}{\@setfontsize\miniscule{5}{6}}
\fi
\makeatother

\DeclareRobustCommand{\optbar}[1]{\shortstack{{\miniscule (\rule[.5ex]{1.25em}{.18mm})}
  \\ [-.7ex] $#1$}}

\def\mumu       {{\ensuremath{\Pmu^+\Pmu^-}}\xspace}

\def\squark    {{\ensuremath{\Ps}}\xspace}

\def\cquark    {{\ensuremath{\Pc}}\xspace}

\def\bquark    {{\ensuremath{\Pb}}\xspace}

\def\pion   {{\ensuremath{\Ppi}}\xspace}

\def\pip    {{\ensuremath{\pion^+}}\xspace}
\def\pim    {{\ensuremath{\pion^-}}\xspace}

\def\kaon    {{\ensuremath{\PK}}\xspace}

\def\KorKbar {\kern \thebaroffset\optbar{\kern -\thebaroffset \PK}{}\xspace}

\def\Kp      {{\ensuremath{\kaon^+}}\xspace}
\def\Km      {{\ensuremath{\kaon^-}}\xspace}

\def\D       {{\ensuremath{\PD}}\xspace}

\def\DorDbar {\kern \thebaroffset\optbar{\kern -\thebaroffset \PD}\xspace}

\def\Dp      {{\ensuremath{\D^+}}\xspace}
\def\Dm      {{\ensuremath{\D^-}}\xspace}

\def\DpDm    {\ensuremath{\Dp {\kern -0.16em \Dm}}\xspace}

\def\Ds      {{\ensuremath{\D^+_\squark}}\xspace}

\def\B       {{\ensuremath{\PB}}\xspace}

\def\BorBbar {\kern \thebaroffset\optbar{\kern -\thebaroffset \PB}\xspace}

\def\Bd      {{\ensuremath{\B^0}}\xspace}

\def\BdorBdbar {\kern \thebaroffset\optbar{\kern -\thebaroffset \Bd}\xspace}
\def\Bu      {{\ensuremath{\B^+}}\xspace}

\def\Bs      {{\ensuremath{\B^0_\squark}}\xspace}

\def\BsorBsbar {\kern \thebaroffset\optbar{\kern -\thebaroffset \Bs}\xspace}

\def\jpsi     {{\ensuremath{{\PJ\mskip -3mu/\mskip -2mu\Ppsi}}}\xspace}
\def\psitwos  {{\ensuremath{\Ppsi{(2S)}}}\xspace}

\def\Y#1S{\ensuremath{\PUpsilon{(#1S)}}\xspace}

\def\proton      {{\ensuremath{\Pp}}\xspace}

\def\Lz          {{\ensuremath{\PLambda}}\xspace}

\def\LorLbar     {\kern \thebaroffset\optbar{\kern -\thebaroffset \PLambda}\xspace}

\def\Xires       {{\ensuremath{\PXi}}\xspace}

\def\Omegares    {{\ensuremath{\POmega}}\xspace}

\def\Lc          {{\ensuremath{\Lz^+_\cquark}}\xspace}

\def\Xicz        {{\ensuremath{\Xires^0_\cquark}}\xspace}
\def\Xicp        {{\ensuremath{\Xires^+_\cquark}}\xspace}

\def\Omegac      {{\ensuremath{\Omegares^0_\cquark}}\xspace}

\def\Xicc        {{\ensuremath{\Xires_{\cquark\cquark}}}\xspace}

\def\Xiccp       {{\ensuremath{\Xires^+_{\cquark\cquark}}}\xspace}
\def\Xiccpp      {{\ensuremath{\Xires^{++}_{\cquark\cquark}}}\xspace}

\def\Omegacc     {{\ensuremath{\Omegares^+_{\cquark\cquark}}}\xspace}

\def\Xiccp       {{\ensuremath{\Xires^+_{\cquark\cquark}}}\xspace}
\def\Xiccpp      {{\ensuremath{\Xires^{++}_{\cquark\cquark}}}\xspace}

\newcommand{\decay}[2]{\ensuremath{\mathinner{#1\!\to #2}}\xspace}

\def\to                 {\ensuremath{\rightarrow}\xspace}

\def\AT#1     {\ensuremath{A_{\mathrm{T}}^{#1}}\xspace}           

\def\C#1      {\ensuremath{\mathcal{C}_{#1}}\xspace}                       
\def\Cp#1     {\ensuremath{\mathcal{C}_{#1}^{'}}\xspace}                    
\def\Ceff#1   {\ensuremath{\mathcal{C}_{#1}^{\mathrm{(eff)}}}\xspace}        
\def\Cpeff#1  {\ensuremath{\mathcal{C}_{#1}^{'\mathrm{(eff)}}}\xspace}       
\def\Ope#1    {\ensuremath{\mathcal{O}_{#1}}\xspace}                       
\def\Opep#1   {\ensuremath{\mathcal{O}_{#1}^{'}}\xspace}                    

\newcommand{\aunit}[1]{\ensuremath{\text{\,#1}}}       

\newcommand{\tev}{\aunit{Te\kern -0.1em V}\xspace}
\newcommand{\gev}{\aunit{Ge\kern -0.1em V}\xspace}
\newcommand{\mev}{\aunit{Me\kern -0.1em V}\xspace}
\newcommand{\kev}{\aunit{ke\kern -0.1em V}\xspace}
\newcommand{\ev}{\aunit{e\kern -0.1em V}\xspace}
 
\newcommand{\mevc}{\ensuremath{\aunit{Me\kern -0.1em V\!/}c}\xspace}
\newcommand{\gevc}{\ensuremath{\aunit{Ge\kern -0.1em V\!/}c}\xspace}
\newcommand{\mevcc}{\ensuremath{\aunit{Me\kern -0.1em V\!/}c^2}\xspace}
\newcommand{\gevcc}{\ensuremath{\aunit{Ge\kern -0.1em V\!/}c^2}\xspace}

\def\cm   {\aunit{cm}\xspace}

\def\fb   {\ensuremath{\aunit{fb}}\xspace}
\def\invfb   {\ensuremath{\fb^{-1}}\xspace}

\def\sec  {\ensuremath{\aunit{s}}\xspace}

\def\ps   {\ensuremath{\aunit{ps}}\xspace}
\def\fs   {\aunit{fs}}

\newcommand{\chisq}{\ensuremath{\chi^2}\xspace}

\newcommand{\chisqip}{\ensuremath{\chi^2_{\text{IP}}}\xspace}

\def\gsim{{~\raise.15em\hbox{$>$}\kern-.85em
          \lower.35em\hbox{$\sim$}~}\xspace}
\def\lsim{{~\raise.15em\hbox{$<$}\kern-.85em
          \lower.35em\hbox{$\sim$}~}\xspace}

\def\pt         {\ensuremath{p_{\mathrm{T}}}\xspace}

\def\mrad{\aunit{mrad}\xspace}

\def\tell1  {TELL1\xspace}
\def\ukl1   {UKL1\xspace}

\newcommand{\lhcborcid}[1]{\href{https://orcid.org/#1}{\hspace*{0.1em}\raisebox{-0.45ex}{\includegraphics[width=1em]{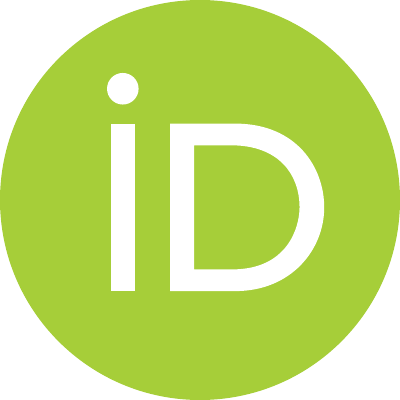}}}}

\def\genxicc     {\mbox{\textsc{GenXicc}}\xspace}

\def\mcand  {\ensuremath{m_{\mathrm{cand}}}\xspace}

\newcommand{\mpdg}[1]{{\ensuremath{M_{\mathrm{PDG}}(#1)}\xspace}}

\mathchardef\mhyphen="2D

\usepackage{cite} 
\usepackage{mciteplus}

\begin{document}

\renewcommand{\thefootnote}{\fnsymbol{footnote}}
\setcounter{footnote}{1}

\begin{titlepage}
\pagenumbering{roman}

\vspace*{-1.5cm}
\centerline{\large EUROPEAN ORGANIZATION FOR NUCLEAR RESEARCH (CERN)}
\vspace*{1.5cm}
\noindent
\begin{tabular*}{\linewidth}{lc@{\extracolsep{\fill}}r@{\extracolsep{0pt}}}
\ifthenelse{\boolean{pdflatex}}
{\vspace*{-1.5cm}\mbox{\!\!\!\includegraphics[width=.14\textwidth]{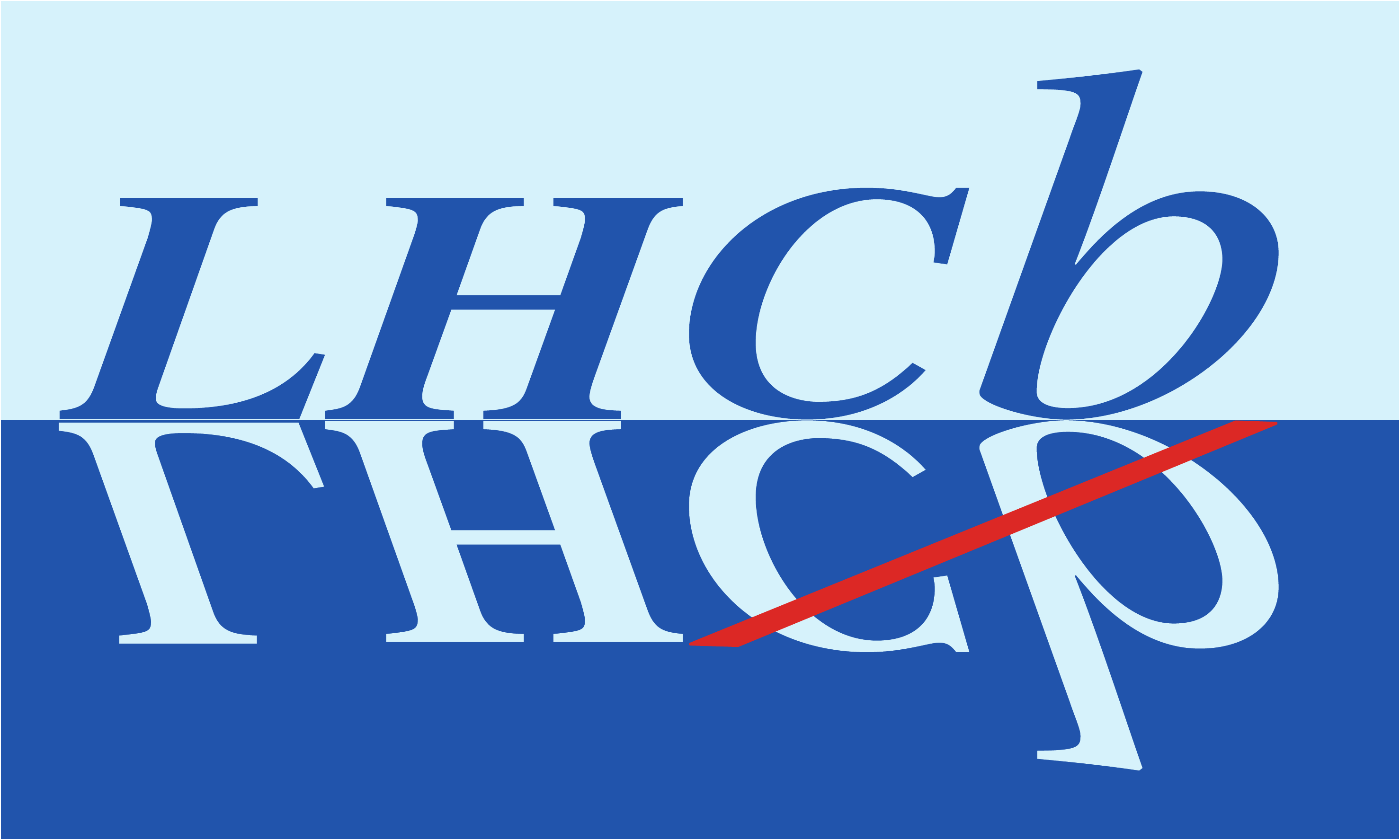}} & &}%
{\vspace*{-1.2cm}\mbox{\!\!\!\includegraphics[width=.12\textwidth]{figs/lhcb-logo.eps}} & &}%
\\
 & & CERN-EP-2026-245 \\  
 & & LHCb-PAPER-2026-022 \\  
 & & September 18, 2026 \\ 
 & & \\
\end{tabular*}

\vspace*{4.0cm}

{\normalfont\bfseries\boldmath\huge
\begin{center}
  \papertitle 
\end{center}
}

\vspace*{2.0cm}

\begin{center}
\paperauthors\footnote{Authors are listed at the end of this Letter.}
\end{center}

\vspace{\fill}

\begin{abstract}
  \noindent
A search for the doubly charmed baryon $\varOmega^+_{cc}$ in the $\varOmega^0_c\pi^+$ decay channel is performed using proton-proton collision data corresponding to an integrated luminosity of $6.3\text{\,fb}^{-1}$, collected with the upgraded LHCb detector in 2024 at a center-of-mass energy of 13.6$\text{\,Te\kern -0.1em V}$.
A peaking structure with a global significance of $8.7\sigma$ is observed in the $\varOmega^0_c\pi^+$ mass spectrum, where the $\varOmega^0_c$ baryon is reconstructed in the $pK^-K^-\pi^+$ final state.
The structure is consistent with originating from a weakly decaying particle and is identified as the doubly charmed baryon $\varOmega^+_{cc}$.
Its mass is determined to be
$3725.9 \pm 1.0 \,(\mathrm{stat}) \pm 0.2 \,(\mathrm{syst}) \pm 0.4 \,(\mathrm{lifetime}) \pm 0.6 \,(\mathrm{ext})\text{\,Me\kern -0.1em V\!/}c^2$,
where the third uncertainty arises from the dependence of the selection-induced bias on the unknown $\varOmega^+_{cc}$ lifetime,
and the fourth is due to the uncertainties on the masses of the $\varOmega^0_c$, $\varXi^+_c$, and $\varXi^{++}_{cc}$ baryons.
\end{abstract}

\vspace*{2.0cm}

\begin{center}
  Submitted to
  Phys.~Rev.~Lett.
\end{center}

\vspace{\fill}

{\footnotesize 
\centerline{\copyright~\papercopyright. \href{\paperlicenceurl}{\paperlicence}.}}
\vspace*{2mm}

\end{titlepage}


\newpage
\setcounter{page}{2}
\mbox{~}


\renewcommand{\thefootnote}{\arabic{footnote}}
\setcounter{footnote}{0}

\cleardoublepage


\pagestyle{plain} 
\setcounter{page}{1}
\pagenumbering{arabic}


\clearpage
The quark model~\cite{GellMann:1964nj,Petermann:1965qlk,Zweig:352337} predicts multiplets of baryon and meson states whose structure is determined by the symmetry properties of the hadron wave functions. 
In the first two generations of quarks---up ($u$), down ($d$), charm ($c$), and strange ($s$)---these states form $\mathrm{SU}(4)_f$ multiplets \cite{su4_1975}. 
All ground states with charm quantum number $C = 0$ or $1$ have been discovered~\cite{PDG2024}. 
Three weakly decaying $C = 2$ states are expected: the \Xicc isodoublet ($\Xiccpp\ (ucc)$ and $\Xiccp\ (dcc)$) and the isosinglet $\Omegacc\ (scc)$.
For the \Omegacc baryon, a variety of phenomenological models and first-principles calculations predict a mass in the range 3660--3780$\mevcc$~\cite{Wang:2010hs,Briceno:2012wt,Durr:2012dw,PACS-CS:2013vie,Padmanath:2013zfa,Brown:2014ena,Padmanath:2015jea,PhysRevD.92.034504,Chen:2017kxr,Alexandrou:2017xwd,Chen:2017sbg,Lu:2017meb,Yao:2018ifh,Weng:2018mmf,Karliner:2018hos,Ma:2017nik,Kerbikov:2018okv,Yu:2018com,Wang:2018lhz,Li:2019ekr,Bahtiyar:2020uuj,Faustov:2020gun,Rahmani:2020pol,Zhang:2021yul,Oudichhya:2022ssc,Tong:2021raz,Ghalenovi:2022dok,Song:2022csw,Yu:2022lel,M:2023hms,Ortiz-Pacheco:2023kjn,He:2023gqh,ShekariTousi:2024mso,Oudichhya:2024hmn,Crede:2024hur,Kosov:2025pnv,Harsha:2025flk,Pan:2025lmd}. 
The lifetime hierarchy of these doubly charmed baryons is expected to be $\tau(\Xiccp) < \tau(\Omegacc) < \tau(\Xiccpp)$,
with recent predictions of $\tau(\Omegacc)$ varying in the range 75--190$\fs$~\cite{Berezhnoy:2018bde,Cheng:2018mwu,Likhoded:2018tfc,Cheng:2026mlv,Dulibic:2026atz}, and Ref.~\cite{Dulibic:2023jeu} predicts the ratio $\tau(\Omegacc)/\tau(\Xiccpp)$ to be $0.52 \pm 0.13^{\:+\:0.03}_{\:-\:0.02}$.

The LHCb collaboration reported the first observation of the $\Xiccpp$ baryon in the $\Lc\Km\pip\pip$ decay channel in 2017~\cite{LHCb-PAPER-2017-018}, using a sample of proton-proton ($pp$) collision data collected at a center-of-mass energy of $\sqrt{s}=13\tev$, corresponding to an integrated luminosity of $1.7\invfb$.
The inclusion of charge-conjugate decay processes is implied throughout this Letter.
Several other \Xiccpp decay channels were later measured by LHCb, including $\Xicp\pip$~\cite{LHCb-PAPER-2018-026}, $\PXi_c^{'+}\pip$~\cite{LHCb-PAPER-2021-052}, and $\Xicz\pip\pip$~\cite{LHCb-PAPER-2024-053},
while no significant signal was found for the $\Dp\proton\Km\pip$ decay channel~\cite{LHCb-PAPER-2019-011}.
The most recent measurements of the mass and lifetime of the \Xiccpp baryon are $3621.55\pm0.23\,(\mathrm{stat})\pm0.30\,(\mathrm{syst})\mevcc$~\cite{LHCb-PAPER-2019-037} and $0.256^{\:+\:0.024}_{\:-\:0.022}\,(\mathrm{stat})\pm0.014\,(\mathrm{syst})\ps$~\cite{LHCb-PAPER-2018-019}. 
Recently, the \Xiccp baryon was observed by the LHCb collaboration in the $\Lc\Km\pip$ decay channel with Run~3 data~\cite{LHCb-PAPER-2026-009},
after several inconclusive searches with LHCb Run~1--2 data~\cite{LHCb-PAPER-2013-049,LHCb-PAPER-2019-029}.
The mass of the \Xiccp baryon was measured to be $3619.97 \pm 0.83\, ({\rm stat})\pm 0.26\,({\rm syst}) ^{\:+\:1.90}_{\:-\:1.30}\, ({\rm lifetime}) \mevcc$, consistent with the interpretation that this identified state is the isospin partner of the \Xiccpp baryon.
A search for the \Omegacc baryon in the $\Xicp\Km\pip$ decay channel was performed by LHCb using $pp$ collision data collected at $\sqrt{s}=13\tev$ corresponding to an integrated luminosity of $5.4\invfb$, but no significant signal was observed~\cite{LHCb-PAPER-2021-011}.

This Letter reports the first observation of the \Omegacc baryon, reconstructed in the \Omegac\pip decay channel, together with the first measurement of its mass.
The choice of this decay mode is motivated by the prediction that  $\Omegacc\to\Omegac\pip$ has one of the largest branching fractions among the hadronic decays of the $\Omegacc$ baryon~\cite{ Wang:2017mqp,Geng:2017mxn,Dhir:2018twm,Zhao:2018mrg,Jiang:2018oak,Gerasimov:2019jwp,Gutsche:2019iac,Cheng:2020wmk,Han:2021azw,Hu:2024uia,Li:2025noz}.
In addition, the \(\Omegac\) lifetime was found to be about four times the previously measured value~\cite{LHCb-PAPER-2018-028,LHCb-PAPER-2021-021,LHCb-PAPER-2025-013}, allowing combinatorial background to be suppressed more effectively by exploiting the displacement of its decay vertex.
The intermediate \Omegac baryon is reconstructed in the $p\Km\Km\pip$ final state.
The measurement uses $pp$ collision data collected by the upgraded LHCb detector in 2024 at $\sqrt{s}=13.6\tev$, corresponding to an integrated luminosity of $6.3\invfb$.
The known $\Xiccpp\to\Xicp\pip$ decay is used as a control channel to validate the selection procedure and reduce systematic uncertainties on the mass measurement, with the \Xicp baryon reconstructed in the $p\Km\pip$ final state.
The relatively high efficiency for reconstructing the intermediate charmed baryons with final-state tracks originating from a common decay vertex partially compensates for the small branching fractions of these Cabibbo-suppressed decay modes, since Cabibbo-favoured \Omegac and \Xicp decays often involve intermediate hyperons, whose cascade decays result in lower reconstruction efficiencies.
To avoid experimenter’s bias, the results of the analysis were not examined until the full procedure had been finalized.

The upgraded LHCb detector~\cite{LHCb-DP-2022-002} is a single-arm forward spectrometer covering the pseudorapidity range $2 < \eta < 5$, designed for
the study of particles containing \bquark\ or \cquark\ quarks. 
It was installed prior to LHC Run~3, to allow effective operation at instantaneous luminosity of about $\mathcal{L}=2\times10^{33}\cm^{-2}\sec^{-1}$, five times that of the Runs 1--2 detector~\cite{LHCb-DP-2014-002,LHCb-DP-2008-001}.
It retains the overall subsystem layout of the original LHCb detector, with major upgrades in most subsystems.
The detector elements that are particularly
relevant to this analysis include: a silicon-pixel vertex detector surrounding the $pp$ interaction region to identify \cquark\ and \bquark\ hadrons from their characteristically long flight distance; 
a tracking system to measure the momentum, $p$, of charged particles;
and two ring-imaging Cherenkov detectors to discriminate between different species of charged hadrons.
The upgraded readout and trigger architecture was designed to exploit the increased computing resources available relative to Runs 1--2, enabling the removal of the hardware trigger stage and the readout of the detector at the full LHC collision frequency~\cite{LHCb-TDR-016,LHCb-TDR-017,LHCb-TDR-018,LHCb-TDR-021}, providing improved trigger efficiency, particularly for hadronic decay channels.
The fully software-based trigger system is implemented in two stages: a GPU-based stage that performs a partial event reconstruction and reduces the data volume
by roughly a factor of 30, followed by a CPU-based stage that provides full offline-quality reconstruction and selection of physics signatures.
In between these two stages, an alignment and calibration of the detector is performed in near real-time and their results are used in the trigger~\cite{LHCb-PROC-2015-011}.
Triggered data further undergo a centralized offline processing step, after which highly automated \lhcb analysis productions deliver physics-analysis-ready data with low latency across the entire \lhcb physics program~\cite{Sprucing, FunTuple}.
For accurate determination of particle masses, the momentum scale is calibrated offline using samples of $\decay{\jpsi}{\mumu}$ and $\decay{\Bu}{\jpsi\Kp}$~decays collected concurrently with the data sample used for this measurement, following a similar procedure of Ref.~\cite{LHCb:2023ood}.
The relative accuracy of the calibration is estimated using samples of fully reconstructed $\PUpsilon$ and \psitwos mesons.

Simulated samples of $\Omegacc$ and $\Xiccpp$ decays are used to optimize event selection and to model the signal invariant-mass distributions.
Signal events are generated with the standard LHCb simulation software~\cite{Sjostrand:2007gs,Lange:2001uf,davidson2015photos,Allison:2006ve, *Agostinelli:2002hh}, using the dedicated \genxicc2.0 generator~\cite{genxicc_2010} for the production of doubly charmed baryons.
In the simulation, the \Omegacc mass and lifetime are set to 3738\mevcc and 160\fs, respectively, while the corresponding \Xiccpp parameters are set to their known values~\cite{PDG2024}. 

The reconstruction and selection of $\Omegacc\to\Omegac\pip$ candidates are performed in several steps.
In the first trigger stage, events are required to contain a particle with a large momentum transverse to the beam direction (\pt) and significant separation from all primary vertices (PVs), or a two-particle vertex displaced from its PV.
The associated PV of a particle is defined as the PV to which it has the smallest impact parameter.
In the second trigger stage, candidate $\Omegac\to p\Km\Km\pip$ decays are reconstructed from four charged particles that form a good-quality vertex and are each inconsistent with originating from any PV.
The \Omegac vertex is required to be displaced from its associated PV,
and a loose upper requirement is imposed on the angle between its momentum and the vector connecting its PV and decay vertex.
Each \Omegac candidate is then combined with an additional charged particle to form an \Omegacc candidate, which is required to have a good-quality vertex located upstream of the \Omegac vertex.
The \Omegacc candidate is required to have $\pt > 3\gevc$ and to be consistent with originating from its PV.
Each of the five final-state particles is required to have a good track-fit quality, particle-identification (PID) information consistent with the appropriate hypothesis ($p$, $K$, or $\pi$), and to pass a minimum \pt threshold in the range 200--500\mevc, depending on the particle species and the data-taking period.
The $\decay{\Xiccpp}{\Xicp\pip}$ control channel is reconstructed and selected similarly, with appropriate adjustments of the thresholds to account for the differences in masses and lifetimes of the charmed baryons.

The dominant background arises from random combinations of tracks and is suppressed offline with a boosted decision tree classifier~(BDT)~\cite{Breiman,AdaBoost}  implemented in the TMVA toolkit~\cite{Hocker:2007ht,*TMVA4}. 
The classifier is trained with simulated $\decay{\Omegacc}{\Omegac\pip}$ decays as the signal proxy and wrong-charge  $\Omegac\pim$ combinations in data as the background proxy. 
For both the signal and background training samples, the mass of candidates is required to fall within a signal search region. This is defined as $[2677,2713]\mevcc$ for the \Omegac candidates, corresponding to approximately $\pm3$ times the \Omegac mass resolution, and $[3660, 3780]\mevcc$ for \Omegacc candidates, which covers the expected mass range according to theoretical predictions.
The BDT exploits differences between signal and background in several quantities, including: the fit qualities of the $\Omegacc$ and $\Omegac$ decay vertices; the \pt and the $\chisqip$, defined as the difference in \chisq of the PV fit with and without the particle in question, of the $\Omegacc$ and $\Omegac$ candidates and their decay products; and the angles between the momenta and the vectors connecting the PV and the decay vertices of the $\Omegacc$ and $\Omegac$ candidates.
It also uses the quality of the kinematic refit of the \Omegacc decay chain~\cite{Hulsbergen:2005pu}, where the \Omegacc candidate is constrained to originate from its PV.
The PID information of each final-state particle is also exploited,
with dedicated calibration data samples used to correct for imperfections in the simulation of the PID variables~\cite{LHCb-DP-2018-001}.
In total, 21 variables are used in the training.
Candidates are retained for analysis only if the BDT response exceeds a threshold optimized by maximizing the figure of merit ${\varepsilon_{\rm sig}}/({n_\sigma/2+\sqrt{N_{\rm bkg}}})$~\cite{Punzi:2003bu}, which is suited to searches for new signals. Here, $\varepsilon_{\rm sig}$ is the signal efficiency evaluated with simulation, and $n_\sigma=5$ corresponds to the target significance of five standard deviations. The background yield $N_{\rm bkg}$ in the \Omegacc signal window of $3720\pm15\mevcc$, corresponding to approximately $\pm2.5$ times the \Omegacc mass resolution, is estimated by counting the yield of the wrong-charge sample in the signal window and scaling it by the yield ratio of $\Omegac\pip$ and $\Omegac\pim$ data samples evaluated in the upper \Omegacc mass sideband region of $[3800, 3870]\mevcc$.
The optimal BDT threshold corresponds to a signal efficiency of approximately 32\% and a background retention of approximately $2.4\times10^{-4}$.
The performance of the BDT classifier is also tested for signal decays simulated under other lifetime hypotheses in the range $90$--$190\fs$.
The classifier trained under the lifetime hypothesis of $160\fs$ shows comparable performance to classifiers optimized for alternative lifetime hypotheses.
A second BDT classifier is trained to select $\Xiccpp\to\Xicp\pip$ decays following the same procedure as for the signal channel. The figure of merit $N^\prime_{\mathrm{sig}}/\sqrt{N^\prime_{\mathrm{sig}}+N^\prime_{\mathrm{bkg}}}$ is used to optimize the BDT threshold to minimize the uncertainty in the \Xiccpp mass peak position.

Additional requirements are applied after the BDT selection to prevent an artificial enhancement of the signal significance.
To reject candidates reconstructed from tracks with shared segments, the angle between each pair of final-state particles is required to be larger than 0.7\mrad.
Background from misidentified $\phi\to\Kp\Km$ decays combined with additional tracks contributes to the \Omegac candidates, while misidentified \mbox{$\Dp\to\Km\pip\pip$}, \mbox{$\Dp\to\Kp\Kp\pim$}, and \mbox{$\Ds\to\Kp\Km\pip$} decays contribute to the \Xicp candidates.
For each background decay, the invariant mass of the relevant track combination is recalculated by assigning each track the mass of the corresponding decay product.
Tighter PID requirements are then applied to candidates whose reassigned masses lie near the known $\phi$, \Dp, or \Ds masses~\cite{PDG2024}. These requirements remove the misidentified backgrounds with negligible loss of signal.
About 2.4\% of the events still contain more than one \Omegacc candidate, in which case one candidate is selected at random and the others are discarded.

\begin{figure}
    \centering
    \includegraphics[width=0.49\linewidth]{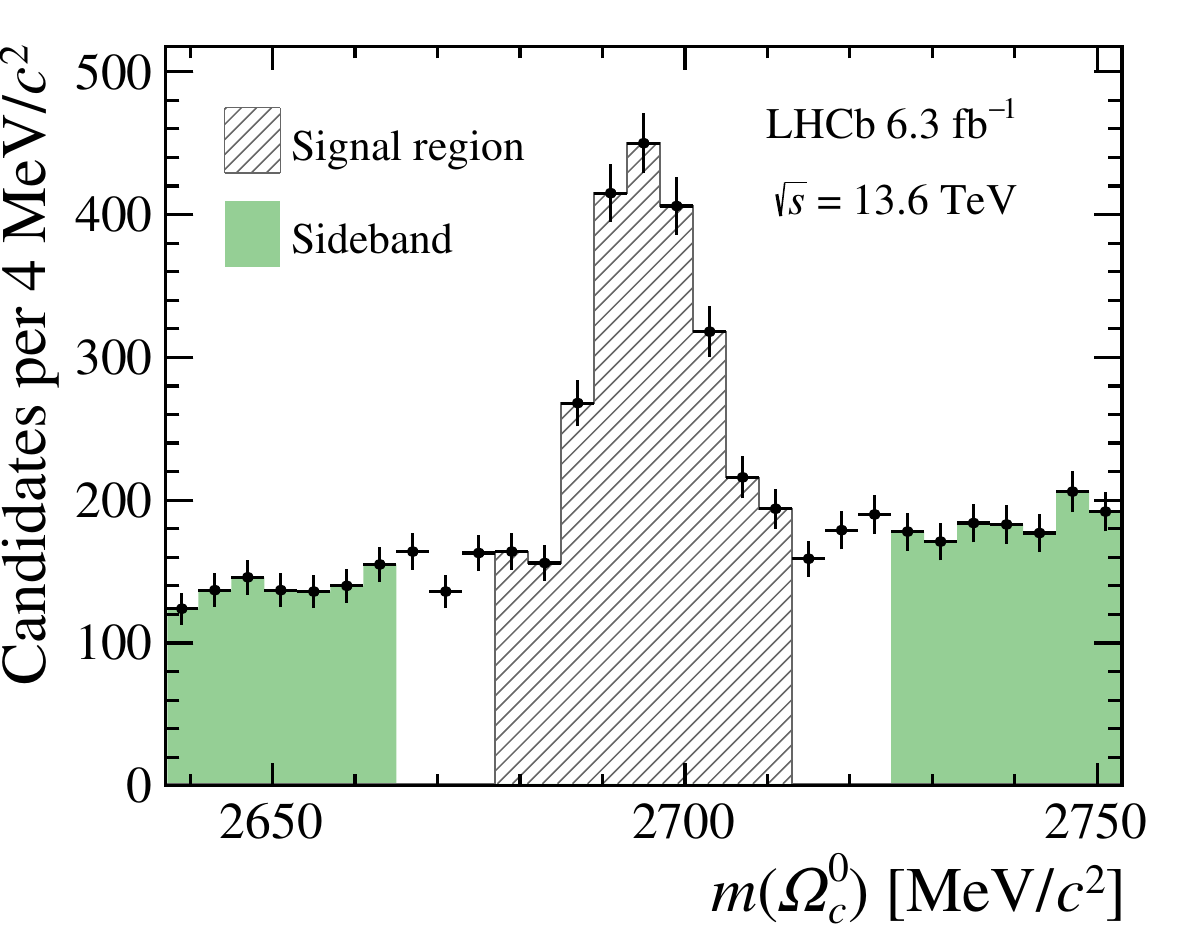}
    \includegraphics[width=0.49\linewidth]{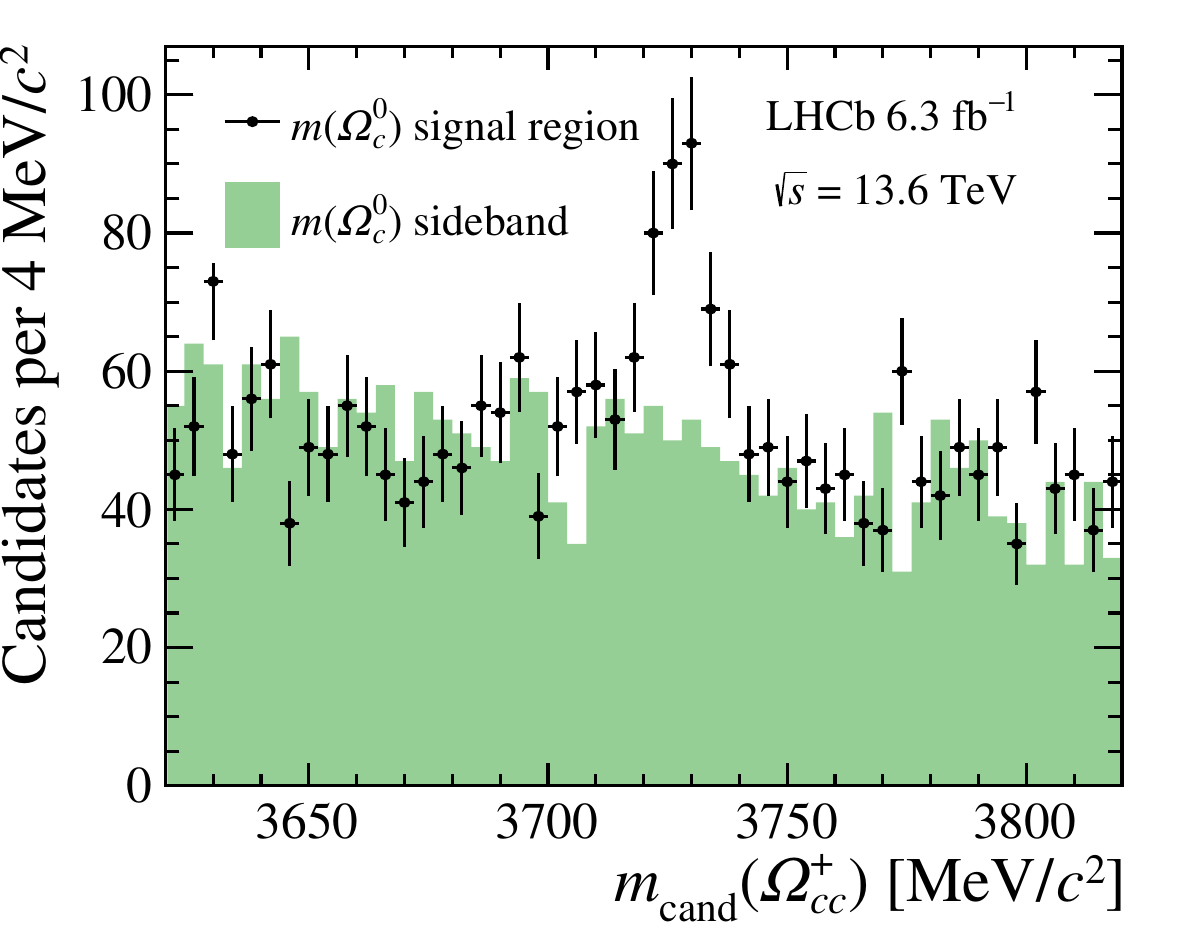}
    \caption{Mass distributions for the (left) intermediate $\varOmega^0_c$ baryon and (right) $\varOmega^+_{cc}$ baryon.
    The $m(\varOmega^0_c)$ signal region is defined as $[2677,2713]\text{\,Me\kern -0.1em V\!/}c^2$, and the sideband region is defined as $[2635,2665]\cup[2725,2755]\text{\,Me\kern -0.1em V\!/}c^2$.}
    \label{fig:mass_after_sel_sig}
\end{figure}

\begin{figure}
    \centering
    \includegraphics[width=0.49\linewidth]{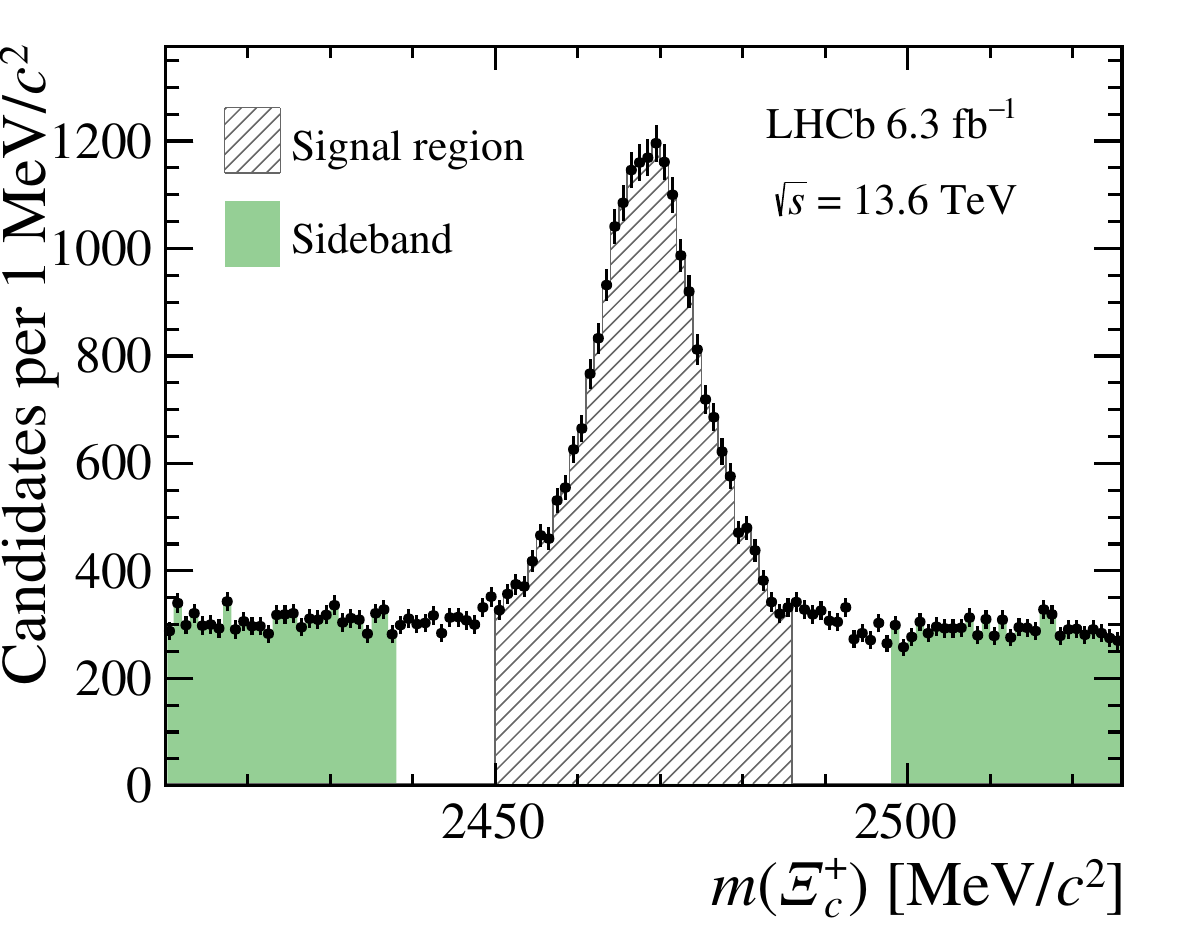}
    \includegraphics[width=0.49\linewidth]{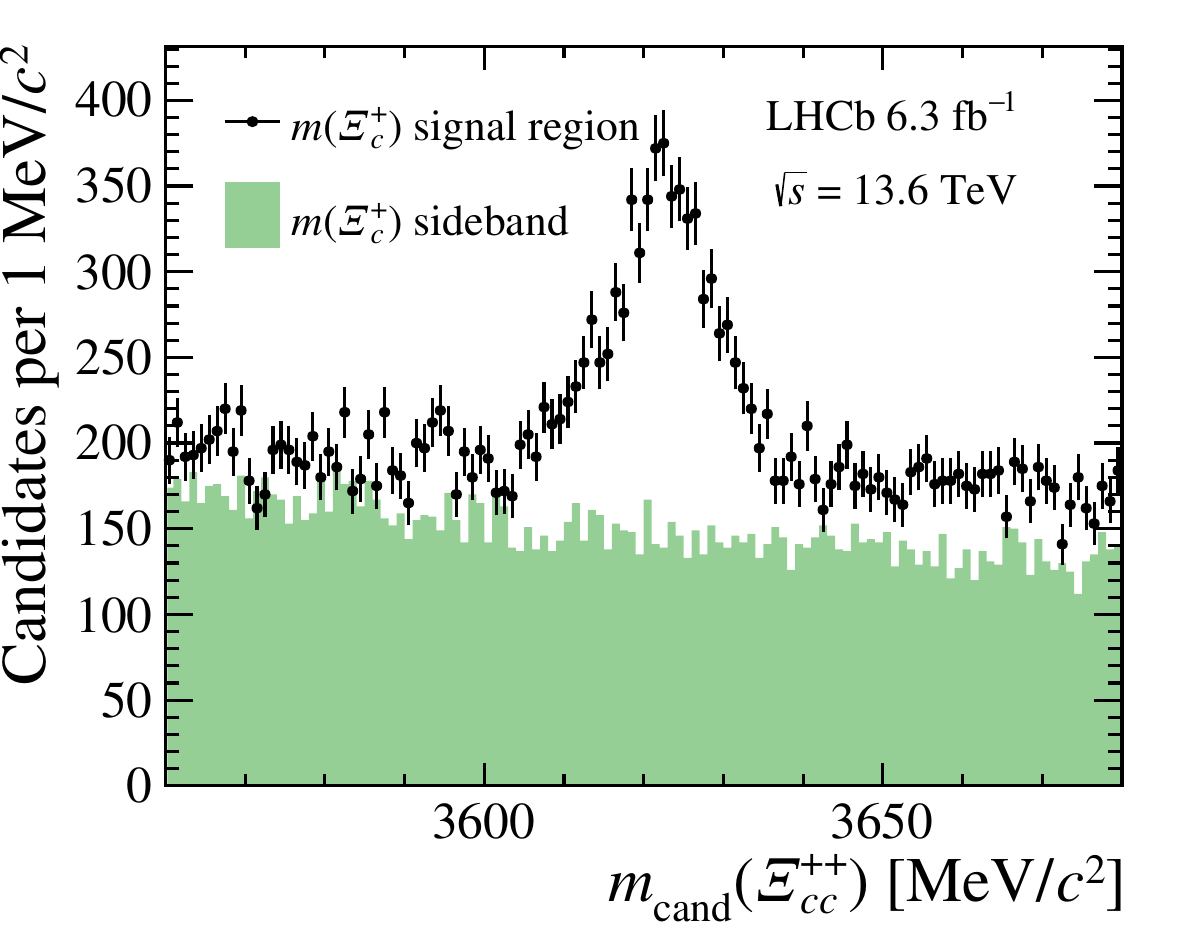}
    \caption{Mass distributions for the (left) intermediate $\varXi^+_c$ baryon and (right) $\varXi^{++}_{cc}$ baryon.
    The $m(\varXi^+_c)$ signal region is defined as $[2450,2486]\text{\,Me\kern -0.1em V\!/}c^2$, and the sideband region is defined as $[2408,2438]\cup[2498,2528]\text{\,Me\kern -0.1em V\!/}c^2$.}
    \label{fig:mass_after_sel_con}
\end{figure}

Mass distributions after the selection are shown in Figs.~\ref{fig:mass_after_sel_sig} and \ref{fig:mass_after_sel_con} for the signal and control channels, respectively, with $\mcand(H_{cc})\equiv m(H_c\pip) - m(H_c) + \mpdg{H_c}$. Here, $H_c$ ($H_{cc}$) denotes the \Omegac (\Omegacc) and \Xicp (\Xiccpp) baryons for the signal and control channels, respectively, $\mcand(H_c\pip)$ and $m(H_c)$ are the reconstructed masses of the $H_c\pip$ combination and the intermediate $H_c$ baryon, respectively, and \mpdg{H_c} is the known $H_c$ mass~\cite{PDG2024}.
Using this mass difference improves the mass resolution and cancels, to leading order, momentum-scale calibration biases associated with the final-state tracks from the intermediate $H_c$ decay.
A peaking structure is observed in the \Omegacc mass spectrum around $3727\mevcc$, while no significant structure is seen for candidates in the \Omegac mass sideband regions.

\begin{figure}
    \centering
    \includegraphics[width=0.49\linewidth]{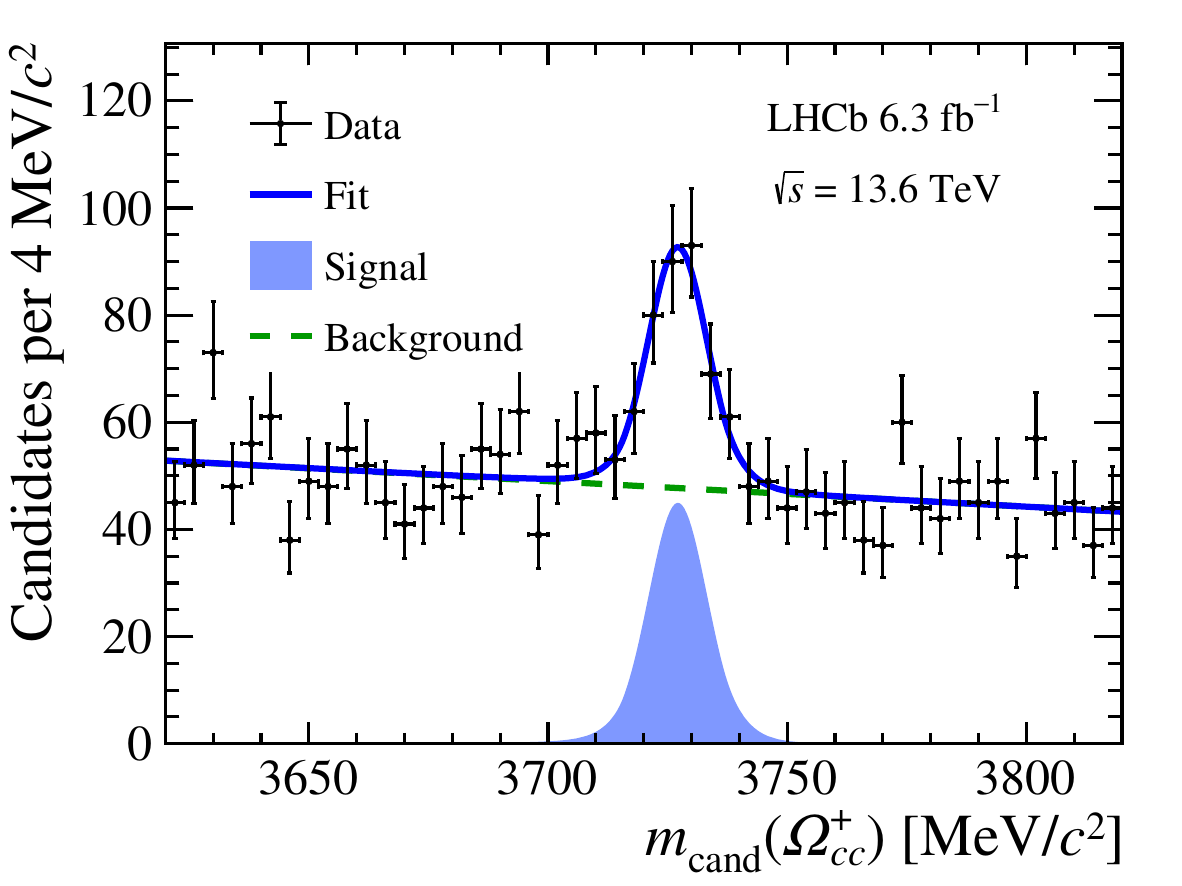}
    \includegraphics[width=0.49\linewidth]{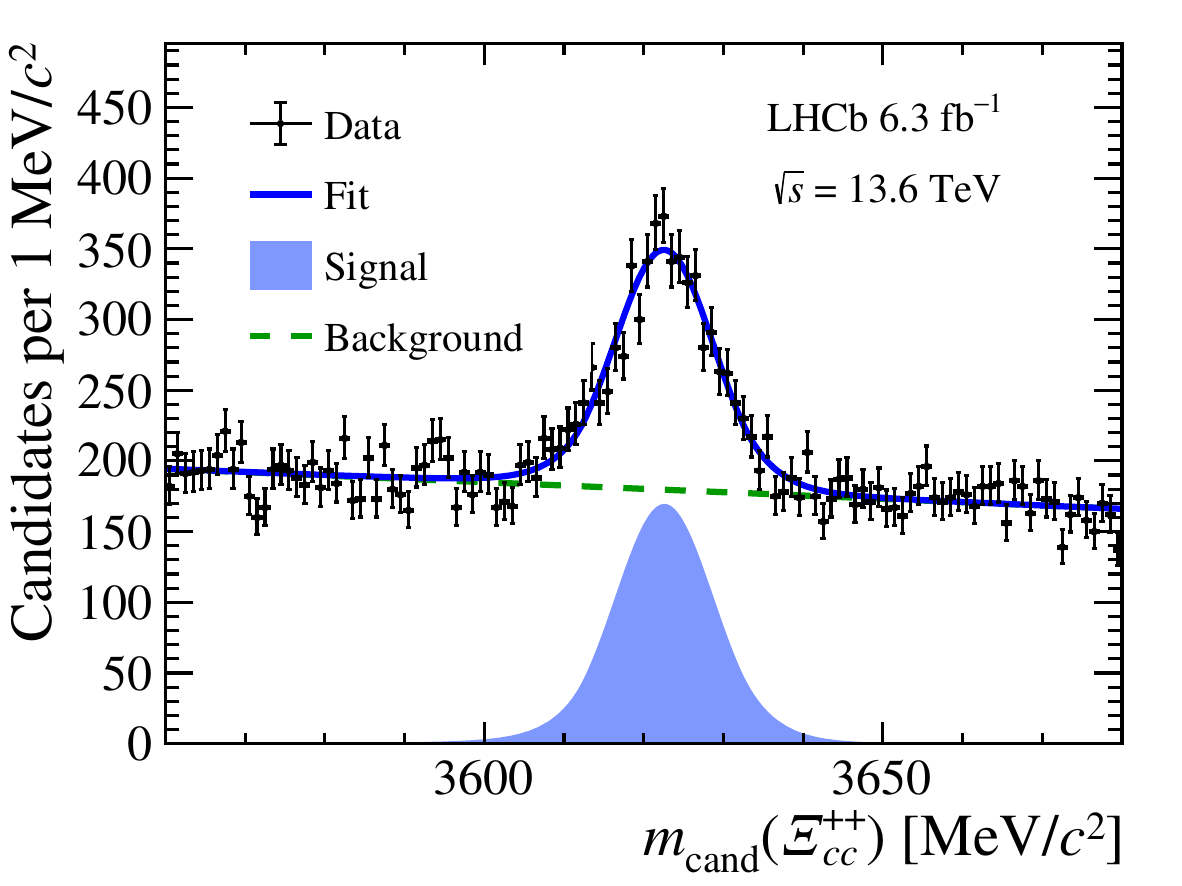}
    \caption{Mass distributions of the (left) $\varOmega^+_{cc}$ and (right) $\varXi^{++}_{cc}$ candidates, with the fit results overlaid.}
    \label{fig:mass_with_fit}
\end{figure}

An unbinned extended maximum-likelihood fit is performed simultaneously on the \Omegacc and \Xiccpp invariant-mass distributions shown in Fig.~\ref{fig:mass_with_fit}.
The signals are described by the sum of a Gaussian function and a modified Gaussian function with power-law tails on both sides~\cite{Skwarnicki:1986xj}.
All shape parameters are fixed to values obtained from simulation, except for the mass peak positions and resolution.
The resolution in the \Xiccpp signal channel is free to vary in the fit, while the resolution in the \Omegacc signal channel is fixed to the simulated value corrected by the difference between data and simulation evaluated with the \Xiccpp sample.
The background is described by a linear function with a slope that is free to vary in the fit.
The \Omegacc signal yield is measured to be $184\pm24$, corresponding to a local significance of $9.0\sigma$ evaluated with a likelihood-ratio test~\cite{significance_2011}.
After accounting for the look elsewhere effect~\cite{Gross:2010qma} in the range 3660--3780\mevcc, the global significance is $8.7\sigma$.
The \Xiccpp signal yield is measured to be $2770\pm110$,
and the \Xiccpp mass resolution is measured to be $6.3\pm0.3\mevcc$, corresponding to an \Omegacc mass resolution of $6.4\mevcc$.
The difference between the mass peak positions of the \Omegacc and \Xiccpp baryons is measured to be $104.6 \pm 1.0\mevcc$.
Using the mass difference cancels, to leading order, the bias associated with the companion pion, as discussed below.

The event selection favors candidates with larger flight distances. In the presence of multiple scattering, this biases the opening angles between the decay products and therefore the reconstructed invariant mass~\cite{LHCb:2023ood}. The size of the effect depends on the selection requirements and on the lifetime of the decaying hadron. 
For weakly decaying charmed hadrons, the bias is typically not negligible. It can be evaluated with simulation by comparing the fitted mass of signal candidates before and after applying the selection~\cite{LHCb-PAPER-2017-018,LHCb-PAPER-2019-037}.
The bias is studied for \Omegacc lifetime hypotheses in the interval $[90,190]\fs$, chosen according to the predicted lifetime ratio $\tau(\Omegacc)/\tau(\Xiccpp)$~\cite{Dulibic:2023jeu} and the measured lifetime of the \Xiccpp baryon~\cite{LHCb-PAPER-2018-019}.
Within this interval, the bias on the mass difference, which is subtracted from the fitted value, is parameterized as $p_0 + p_1 \times \tau(\Omegacc)$, where \mbox{$p_0 = 1.3\pm0.1\mevcc$} and \mbox{$p_1 = (-7.4\pm0.7)\times10^{-3}\mevcc\,\mathrm{fs}^{-1}$}.
At the midpoint, $\tau(\Omegacc)=140\fs$, the correction is $0.30\mevcc$.

The main sources of systematic uncertainty on the measurement of the mass difference are summarized in Table~\ref{tab:syst}, including those associated with the selection-induced bias correction, the unknown \Omegacc lifetime, and the uncertainties on the masses of the charmed baryons.
The dependence of the selection-induced bias correction on the \Omegacc lifetime is modeled with a linear function. The uncertainty associated with the parametrization of this lifetime dependence is evaluated using an exponential function as an alternative model.
The uncertainty due to the limited knowledge of the \Xiccpp lifetime is evaluated by varying the lifetime hypothesis used for the simulated \Xiccpp decays.
The correction is evaluated using simulation samples, and the uncertainty due to their limited size is propagated.
The combined uncertainty amounts to 0.2\mevcc and is assigned as the uncertainty on the selection-induced bias correction.
The maximum deviation of the correction from its value at the baseline lifetime of $140\fs$ over the range $[90,190]\fs$ is 0.37\mevcc, which is assigned as the systematic uncertainty associated with the unknown \Omegacc lifetime.
Since the definition of $m_{\mathrm{cand}}$ involves the known masses of the \Omegac and \Xicp baryons, their uncertainties of 0.41\mevcc and 0.23\mevcc~\cite{PDG2024,LHCb-PAPER-2014-021,LHCb-PAPER-2023-011}, respectively, are propagated to the measured mass difference.
The uncertainty on the \Xiccpp mass is additionally propagated when converting the mass difference to the \Omegacc mass.
Other sources of systematic uncertainty are evaluated and found to have effects below $0.1\mevcc$ on the measured mass difference and are considered negligible.
The effect of the imperfect knowledge of the momentum scale is evaluated by varying the calibration constants within their uncertainties and determining the change in the measured mass difference.
The momenta of the final-state particles are corrected for ionization energy loss in the detector material.
The residual bias in this correction is determined from \decay{\psitwos}{\jpsi\pip\pim} decays~\cite{Chu:2025qyv}.
This bias is propagated through the decay kinematics and found to induce a small shift in the measured mass difference.
The systematic uncertainty associated with the fit model is estimated using alternative signal (a sum of two Gaussian functions) and background (an exponential function) models that describe the data equally well. 
The corresponding biases are estimated using pseudoexperiments in which samples are generated with the default model and fitted with each of the two alternative models. 
\begin{table}
  \centering
  \caption{Sources of systematic uncertainty on the mass-difference measurement.}
  \label{tab:syst}
  \begin{tabular}{lc}
    \hline
    Source    &  Size [\mevcc] \\
    \hline
    Selection-induced bias correction    & 0.2 \\
    Unknown \Omegacc lifetime            & 0.4 \\
    Uncertainty on charmed-baryon masses & 0.5 \\
    \hline
    Statistical uncertainty              & 1.0 \\ 
    \hline
  \end{tabular}
\end{table}

Additional studies are performed to test the robustness of the observation.
The local significance of the structure in the $\Omegac\pip$ mass spectrum remains above $9\sigma$ when candidates are required to have a decay time larger than 170\fs, approximately three times the decay-time resolution.
The persistence of the structure with this minimal decay-time threshold is consistent with the hypothesis that it arises from a weakly decaying state, and disfavors an interpretation as a strongly decaying resonance.
The mass difference is found to be consistent across subsamples obtained by splitting the data according to data-taking period, magnet polarity, and particle or antiparticle decay.

In summary, a significant peaking structure is observed in the $\Omegac\pip$ mass spectrum using a $pp$ collision data sample collected by the upgraded LHCb detector at \mbox{$\sqrt{s}=13.6\tev$}.
The observed width is consistent with the experimental resolution, and the signal candidates exhibit significant decay lengths.
The state is therefore incompatible with a strongly decaying particle and is identified as the doubly charmed baryon \Omegacc.
The mass difference between the \Omegacc and \Xiccpp baryons is determined to be 
\begin{equation*}
M(\Omegacc)-M(\Xiccpp)=104.3 \pm 1.0 \,(\mathrm{stat}) \pm 0.2 \,(\mathrm{syst}) \pm 0.4 \,(\mathrm{lifetime}) \pm 0.5 \,(\mathrm{ext})\mevcc,
\end{equation*}
where the third uncertainty arises from the dependence of the selection-induced bias on the unknown \Omegacc lifetime, 
and the fourth is due to the limited knowledge of masses of the \Omegac and \Xicp baryons.
Using the known \Xiccpp mass~\cite{LHCb-PAPER-2019-037} as an external input, the mass of the \Omegacc baryon is then determined to be 
\begin{equation*}
M(\Omegacc)=3725.9 \pm 1.0 \,(\mathrm{stat}) \pm 0.2 \,(\mathrm{syst}) \pm 0.4 \,(\mathrm{lifetime}) \pm 0.6 \,(\mathrm{ext})\mevcc.
\end{equation*}
With this observation, all members of the $\mathrm{SU}(4)_f$ spin-$\frac{1}{2}$ 20-plet of ground-state baryons predicted by the quark model have now been established experimentally.
Future studies using this dataset, together with larger samples already collected by the upgraded LHCb detector, will probe further properties of the \Omegacc baryon, including its lifetime and other decay channels.

\section*{Acknowledgements}
%
%
\noindent We express our gratitude to our colleagues in the CERN
accelerator departments for the excellent performance of the LHC. We
thank the technical and administrative staff at the LHCb
institutes.
We acknowledge support from CERN and from the national agencies:
ARC (Australia);
CAPES, CNPq, FAPERJ and FINEP (Brazil); 
MOST and NSFC (China); 
CNRS/IN2P3 and CEA (France);  
BMFTR, DFG and MPG (Germany);
NKFIH (Hungary);              
INFN (Italy); 
NWO (Netherlands); 
MNiSW and NCN (Poland); 
MEC/IFA (Romania); 
MICIU and AEI (Spain);
SNSF and SER (Switzerland); 
NASU (Ukraine); 
STFC (United Kingdom); 
DOE NP and NSF (USA).
We acknowledge the computing resources that are provided by ARDC (Australia), 
CBPF (Brazil),
CERN, 
IHEP and LZU (China),
IN2P3 (France), 
KIT and DESY (Germany), 
INFN (Italy), 
SURF (Netherlands),
Polish WLCG (Poland),
IFIN-HH (Romania), 
PIC (Spain), CSCS (Switzerland), 
GridPP (United Kingdom),
and NSF (USA).  
We are indebted to the communities behind the multiple open-source
software packages on which we depend.
Individual groups or members have received support from
RTP (Australia), 
FWO Odysseus grant G0ASD25N (Belgium), 
Key Research Program of Frontier Sciences of CAS, CAS PIFI, CAS CCEPP (China); 
Minciencias (Colombia);
EPLANET, Marie Sk\l{}odowska-Curie Actions, ERC and NextGenerationEU (European Union);
A*MIDEX, ANR, IPhU and Labex P2IO, and R\'{e}gion Auvergne-Rh\^{o}ne-Alpes (France);
Alexander-von-Humboldt Foundation (Germany);
ICSC (Italy); 
Severo Ochoa and Mar\'ia de Maeztu Units of Excellence, GVA, XuntaGal, GENCAT, InTalent-Inditex and Prog.~Atracci\'on Talento CM (Spain);
the Leverhulme Trust, the Royal Society and UKRI (United Kingdom).

\clearpage
\addcontentsline{toc}{section}{References}
\bibliographystyle{LHCb}
\bibliography{main,standard,LHCb-PAPER,LHCb-TDR,LHCb-DP}

\clearpage
\centerline
{\large\bf LHCb collaboration}
\begin
{flushleft}
\small
R.~Aaij$^{39}$\lhcborcid{0000-0003-0533-1952},
M.~Abdelfatah$^{71}$,
A.S.W.~Abdelmotteleb$^{59}$\lhcborcid{0000-0001-7905-0542},
C.~Abellan~Beteta$^{53}$\lhcborcid{0009-0009-0869-6798},
F.~Abudin\'en$^{61}$\lhcborcid{0000-0002-6737-3528},
T.~Ackernley$^{63}$\lhcborcid{0000-0002-5951-3498},
A.A.~Adefisoye$^{71}$\lhcborcid{0000-0003-2448-1550},
B.~Adeva$^{49}$\lhcborcid{0000-0001-9756-3712},
M.~Adinolfi$^{57}$\lhcborcid{0000-0002-1326-1264},
P.~Adlarson$^{87,44}$\lhcborcid{0000-0001-6280-3851}
% ...

{\footnotesize \it

% ...
$^{39}$Nikhef National Institute for Subatomic Physics, Amsterdam, Netherlands\\
% ...
$^{44}$National Center for Nuclear Research (NCBJ), Warsaw, Poland\\
% ...
$^{49}$Instituto Galego de F{\'\i}sica de Altas Enerx{\'\i}as (IGFAE), Universidade de Santiago de Compostela, Santiago de Compostela, Spain\\
% ...
$^{53}$Physik-Institut, Universit{\"a}t Z{\"u}rich, Z{\"u}rich, Switzerland\\
% ...
$^{57}$H.H. Wills Physics Laboratory, University of Bristol, Bristol, United Kingdom\\
% ...
$^{59}$Department of Physics, University of Warwick, Coventry, United Kingdom\\
% ...
$^{61}$School of Physics and Astronomy, University of Edinburgh, Edinburgh, United Kingdom\\
$^{62}$School of Physics and Astronomy, University of Glasgow, Glasgow, United Kingdom\\
$^{63}$Oliver Lodge Laboratory, University of Liverpool, Liverpool, United Kingdom\\
% ...
$^{71}$Syracuse University, Syracuse, NY, United States\\
% ...
$^{87}$Department of Physics and Astronomy, Uppsala University, Uppsala, Sweden, associated to $^{62}$\\
% ...
}
\end{flushleft}



 
\end{document}